\documentclass{article} 

\usepackage{graphicx} 
\usepackage{xcolor} 
\usepackage{amsmath} 
\usepackage{mhchem} 
\usepackage{authblk} 
\usepackage{soul}

\title{Influenza A Virus as a Multivalent Active Particle: Statistical Mechanics of Binding, Motility, and Antiviral Design}

\author[1]{Sumati Bhatia} 
\author[2]{Andreas Herrmann}
\author[3]{Thomas Kolbe} 
\author[1]{Lalan Kumar} 
\author[4]{Daniel Lauster} 
\author[5]{Susanne Liese} 
\author[3]{Bortolo Matteo Mognetti} 
\affil[1]{Department of Chemistry, Faculty of Science and Engineering, Swansea University, Swansea SA2 8PP, United Kingdom\\ \texttt{sumati.bhatia@swansea.ac.uk}} 
\affil[2]{Institut für Chemie und Biochemie, SupraFAB, Freie Universität Berlin, Altensteinstraße 23a, D-14195 Berlin, Germany\\
\texttt{a.herrmann2@fu-berlin.de}}
\affil[3]{Department of Physics, Université libre de Bruxelles, Campus Plaine 1050, Brussels, Belgium\\ \texttt{bortolo.matteo.mognetti@ulb.be}} 
\affil[4]{
Institute of Pharmacy, Biopharmaceuticals, Freie Universität Berlin, Germany\\
\texttt{daniel.lauster@fu-berlin.de}
}
\affil[5]{
Faculty of Mathematics, Natural Sciences, and Materials Engineering, Institute of Physics, University of Augsburg, Augsburg, Germany\\
\texttt{susanne.liese@physik.uni-augsburg.de}
}

\date{\today}

\begin{document}

\maketitle

\begin{abstract}

In this contribution, we review existing multivalent models of Influenza A Virus (IAV) systems to study virion dynamics and design of multivalent entry blockers. We review existing biochemical estimations of the dissociation constants and catalytic activities of the two virion ligands [hemagglutinin (HA) and neuraminidase (NA)], which can be used to develop quantitative theories and experiments. Multivalent models are statistical mechanics frameworks for studying the simultaneous interaction of multiple reversible ligand-receptor bonds. They quantify the effects of both combinatorial entropy terms and single-molecule ligand-receptor affinities. Multivalent frameworks led to the rationale of superselectivity, the ability of probes to sense specific concentrations of conjugated receptors. Multivalent models guided the design of antiviral nanomedicines constituted by polymeric scaffolds conjugated with small molecules such as sialic acid receptors targeting the ligands on the surface of the virion. Importantly, neuraminidase is an enzyme that limits the number of host-guest bonds through receptor desialylation. Like active particles, neuraminidase activity enhances virion motility in the extracellular medium. Modeling the dynamics of Influenza A virions requires developing out-of-equilibrium multivalent models.

\end{abstract}
\begin{center} {Authors listed in alphabetical order} \end{center}

\section{Influenza A Virus Systems}

Influenza A virus (IAV) is a major respiratory pathogen that infects humans and a wide range of animal species, including birds and other mammals \cite{dou2018influenza}. IAV particles are pleomorphic \cite{harris2006influenza,nayak2009influenza,badham2016filamentous}, enveloped virions containing a genome composed of eight segments of negative-sense single-stranded RNA (-ssRNA) \cite{dou2018influenza,hepp2025high}. Virions often carry an incomplete set of genetic segments \cite{jacobs2019incomplete,ganti2022influenza}.

\begin{figure}
    \centering
    \includegraphics[width=0.95\linewidth]{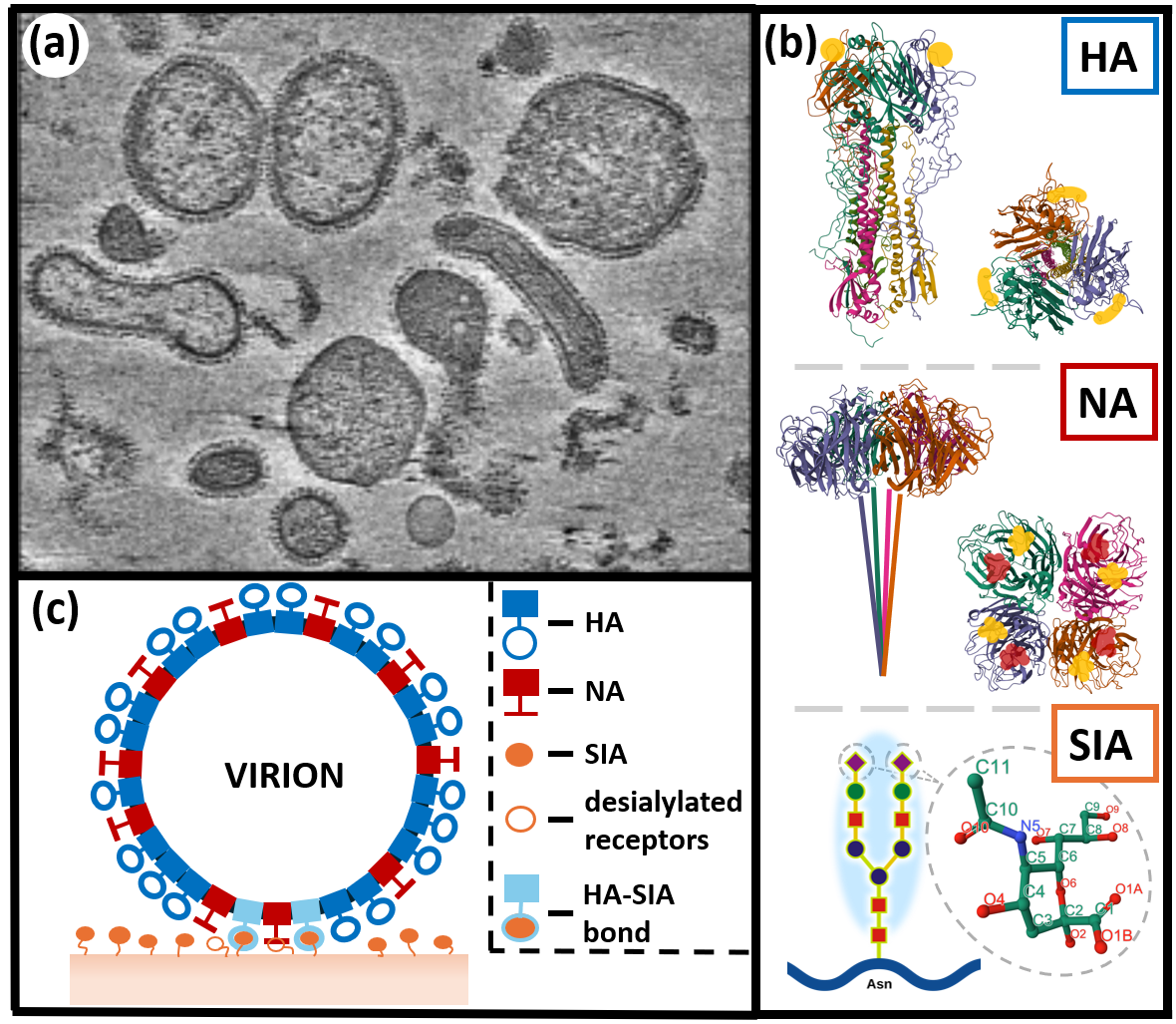}
    \caption{(a) Cryo-electron tomogram of Influenza A/PR8 virions featuring different morphologies. Adapted from \cite{nayak2009influenza}, Copyright 2009, with permission from Elsevier. (b) Side and top views of the crystal structures of the HA trimer (top panel, PDB: 1HGG \cite{PDB_1HGG}) and NA tetramer (middle panel, PDB: 2HTY \cite{PDB_2HTY}). The structure of the NA stalk domain has not been resolved and is therefore schematically depicted as rods. Sialic acid (SIA)-binding sites in HA and secondary binding sites in NA are highlighted in yellow, while the catalytic sites in NA are shown in red, based on \cite{Sung2010}. Bottom panel: Exemplary schematic of sialic acids (purple diamonds) capping an N-linked glycosylated protein at asparagine (Asn) residues, adapted from \cite{KosigrimWiki}. The inset shows the chemical structure of sialic acid (image from the RCSB PDB, created with {Mol*} \cite{Mol_software}). (c) Cartoon illustration of the interaction between HA and NA carried by an Influenza virus particle and SIA distributed over a surface. 
    }
    \label{Fig:General_Model}
\end{figure}

IAV virions are shed into the extracellular environment through budding. IAV particles occur predominantly as spherical, with an $\approx 120\,$nm diameter, or filamentous virions, with a length ranging from a few hundreds of nanometers to a few micrometers (Fig.~\ref{Fig:General_Model}a). Such a classification is approximate as virions with less defined shapes have been reported \cite{harris2006influenza}.
Filamentous virions are more frequently reported {\em in vitro} than {\em in vivo} conditions. The two types of virions also respond differently (e.g., when inhibited with antibodies) during infection progression \cite{peterl2025morphology}.

Two genetic segments (segment 4 and 6) encode for the two spike proteins, hemagglutinin (HA) and neuraminidase (NA), found at the surface of the virions \cite{benton_biophysical_2015,reiter-scherer_force_2019,sung_role_2010} (Fig.~\ref{Fig:General_Model}b-c). HA and NA are key regulators of IAV virion dynamics. HA ligands bind sialic acid (SIA) groups terminating receptor molecules (e.g., glycolipids, Fig.~\ref{Fig:General_Model}b) and macromolecules in the extracellular environment. In contrast, NA molecules are enzymes that cleave SIA groups from their backbones. The catalytic activity of NA reduces the number of SIA groups in the virion's environment, and therefore the number of SIA-HA bonds \cite{macauslane2026neuraminidase}.

There are 18 HA and 11 NA subtypes (noted with HA1-18 and NA1-11), characterized by different affinities with SIA and the type of covalent bond between SIA and its supporting scaffold (with avian and human virions preferentially targeting linkages of type $\alpha$2-3 and $\alpha$3-6, respectively). Due to its essential role in viral attachment, the sialic acid binding site of hemagglutinin is functionally constrained and therefore comparatively conserved, making it an attractive target for antiviral intervention. However, this shallow, non-enzymatic binding site is difficult to address with conventional small-molecule inhibitors. Multivalent approaches can overcome this limitation by presenting multiple sialoside ligands on polymeric or supramolecular scaffolds, thereby increasing avidity toward hemagglutinin and efficiently competing with host-cell receptors.

Single SIA-HA bonds are weak (within the millimolar range \cite{sauter_hemagglutinins_1989}) and reversible. However, due to the high abundance of HA on the virus envelope ($\sim$300 proteins {\em per} spherical particle) and a high SIA density on a mammalian cell surface ($\rho_{\it SIA}\approx 0.2-2/\mathrm{nm}^{2}$ \cite{seaman1977red}), multivalent binding can occur and increase affinity to the picomolar range \cite{benton_biophysical_2015}.
Decoy polymers bearing SIA have been demonstrated to be efficient antivirals, but also tools to investigate virus-host cell interaction. The aim of this work is twofold. First, we review biophysical models of virion dynamics at the cell surface (crawling and detachment) and in mucus. Secondly, we discuss current antiviral designs of functionalized polymers targeting IAV virions. These two types of dynamic interactions are studied using multivalent models \cite{bell_models_1978,martinez-veracoechea_designing_2011,varilly2012general,angioletti2017theory}.

Most of the existing multivalent frameworks are equilibrium models. IAV systems are more peculiar because of the activity of NA enzymes, which steadily decrease the number of available SIA groups in the virion's environment. Here, we discuss the peculiarities of the virion dynamics arising from the action of NA. NA activity facilitates IAV movement across the cell surface by cleaving sialic acid residues that would otherwise restrict virion mobility through repeated interactions with hemagglutinin.
Beyond its established role in virion release from the host-cell membrane, NA may also facilitate entry by helping virions traverse mucus \cite{cohen_influenza_2013,yang2014beneficiary,wang2016influenza,kaler_influenza_2022,kaler_mucus_2025,liu2026filamentous}. In this respect, IAV virions can be considered as active particles \cite{bechinger2016active} powered by desialylation reactions.

Understanding the dynamics of IAV particles is a multiscale effort. Therefore, this work is divided into several sections representing different aspects of the virus-host cell interaction. In Sec.\ \ref{Sec:LR}, we briefly describe the interaction between HA and NA molecules with SIA at the single-molecule level. We discuss biochemical literature reporting ligand-receptor activity. In Sec.\ \ref{Sec:Multivalent}, we review multivalent theories (accounting for combinatorial terms) and superselectivity \cite{martinez-veracoechea_designing_2011}. 
In Sec.~\ref{Sec:Nanom}, we rationalize recent antiviral designs based on multivalency. 
Finally, in Sec.~\ref{Sec:Lim_Multi} we highlight the necessity of considering NA activity and non-equilibrium frameworks to model extracellular dynamics.
Finally, in Sec.~\ref{Sec:Perspective} we present our conclusions and list open challenges. 

\begin{figure}
    \centering
    \includegraphics[width=0.95\linewidth]{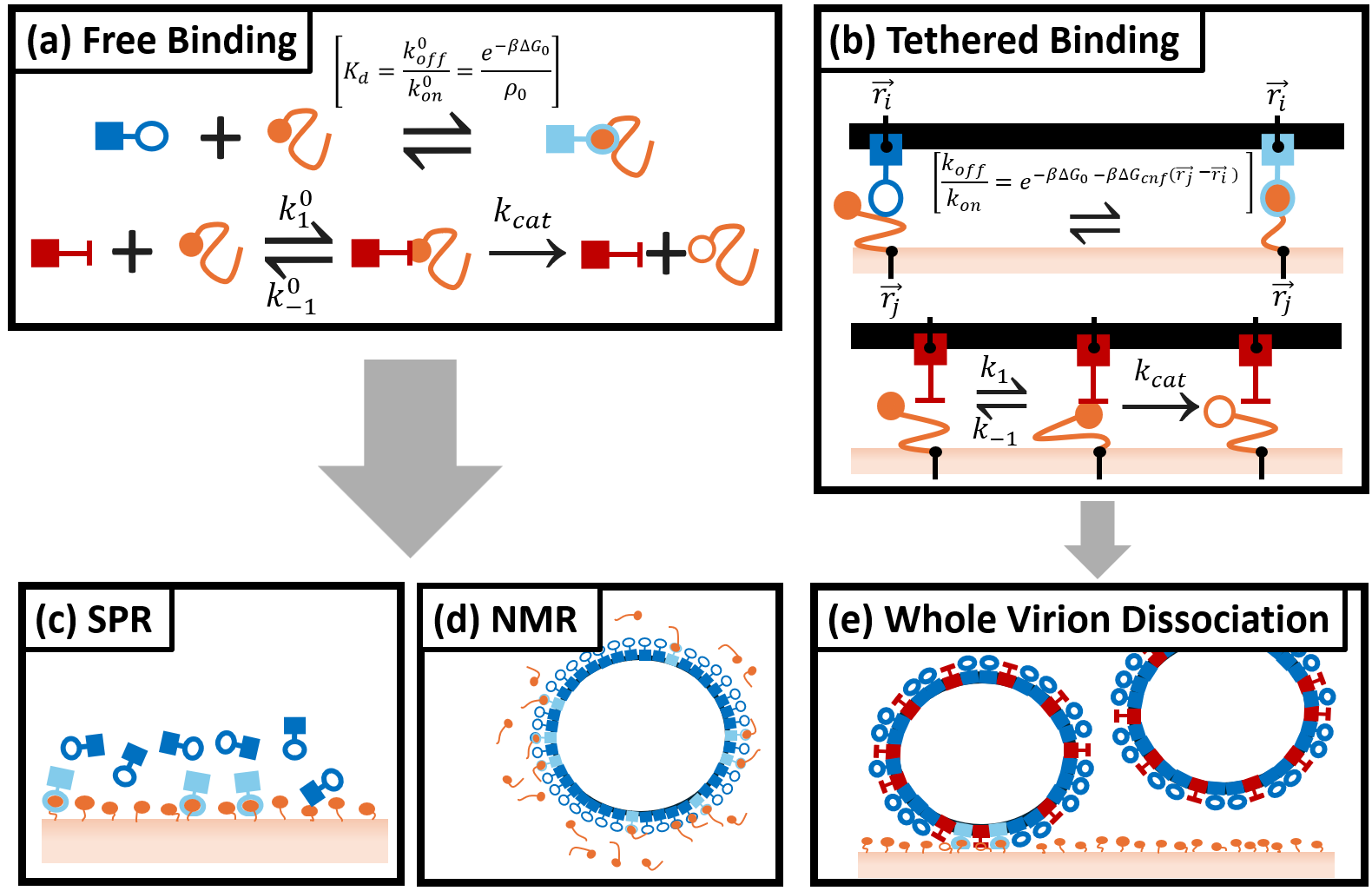}
    \caption{
    (a,b) Schematic contrasting the binding reaction between conjugated ligand-receptor molecules when free in bulk (a) and when tethered to surfaces (b). In the latter case, the dissociation constant includes a configurational term contribution ($\Delta G_{\it cnf}$) that accounts for the reduction in the number of possible scaffold configurations following bond formation. Note that $k_{\it off}$ and $k_{\it on}$ (similarly $k_1$, $k_{-1}$) are defined at the single-molecule level. Instead, for molecules in bulk, the dissociation constant is measured in terms of the equilibrium concentrations of the reactants and products. 
    SPR, force microscopy, and NMR typically measure bulk reaction rates (c,d). Instead, apparent avidity is controlled by the binding rates of surface-bound ligands/receptors (e).  
    }
    \label{Fig:Bond Interaction}
\end{figure}

\section{Single-Molecule Ligand–Receptor Interactions}\label{Sec:LR}

Hemagglutinin is a trimer that binds to sialic acid  residues, which are abundantly found on the surface of epithelial cells and in the protective mucus layer. The principal SIA binding site is placed at the distal tip of each subunit (Fig.~\ref{Fig:General_Model}b). The existence and function of a second binding site are still debated (Sec.~\ref{Sec:second_HA_site}).
The second spike protein on the viral envelope, neuraminidase, is a tetramer that catalyzes the removal of SIA groups from their scaffold (desialylation) \cite{benton_biophysical_2015,reiter-scherer_force_2019,sung_role_2010}. Depending on the virus strain, NA can feature a second functional binding site (Sec.~\ref{Sec:second_HA_site}).

From a biochemical perspective, the interaction between HA and SIA is characterized by the dissociation constant $K_d$ (or the equilibrium constant $K_{\it eq}$, $K_{\it eq}=1/K_{\it d}$) and the binding rates  (Fig.~\ref{Fig:Bond Interaction}a), $k^0_{\it on}$ and $k^0_{\it off}$, with $K_d = k^0_{\it off}/k^0_{\it on}$. 
Similarly, the action of NA is characterized by {\it on}/{\it off} rates, describing the formation of a NA-SIA complex ($k^0_{1}$ and $k^0_{-1}$), and the rate $k_{\it cat}$ at which a SIA residue is cleaved from its scaffold. 
Using the Michaelis–Menten model, we have
\begin{eqnarray}
\ce{NA + SIA <=>[$k^0_1$][$k^0_{-1}$] NA\mathrm{-}SIA ->[$k_{\it cat}$] NA + SIA^*}
\label{Eq:MM}
\end{eqnarray}
where SIA$^*$ schematically denotes the desialylated scaffold and the cleaved SIA. $K_m$ is the Michaelis constant, quantifying the substrate concentration at which the reaction velocity $v$ is half its maximum value ($v_{\it max}=\rho_{\it NA} k_{\it cat}$, where $\rho_{\it NA}$ is NA density)
\begin{eqnarray}
v &=& {v_{\it max} \rho_{\it SIA}\over K_m + \rho_{\it SIA}}
\quad ,
\nonumber
\end{eqnarray}
where $\rho_{\it SIA}$ is SIA density. The binding and cleaving are operated by two different domains present on each of the four NA units  (Fig.~\ref{Fig:General_Model}b).

\begin{figure}
    \centering
    \includegraphics[width=1.0\linewidth]{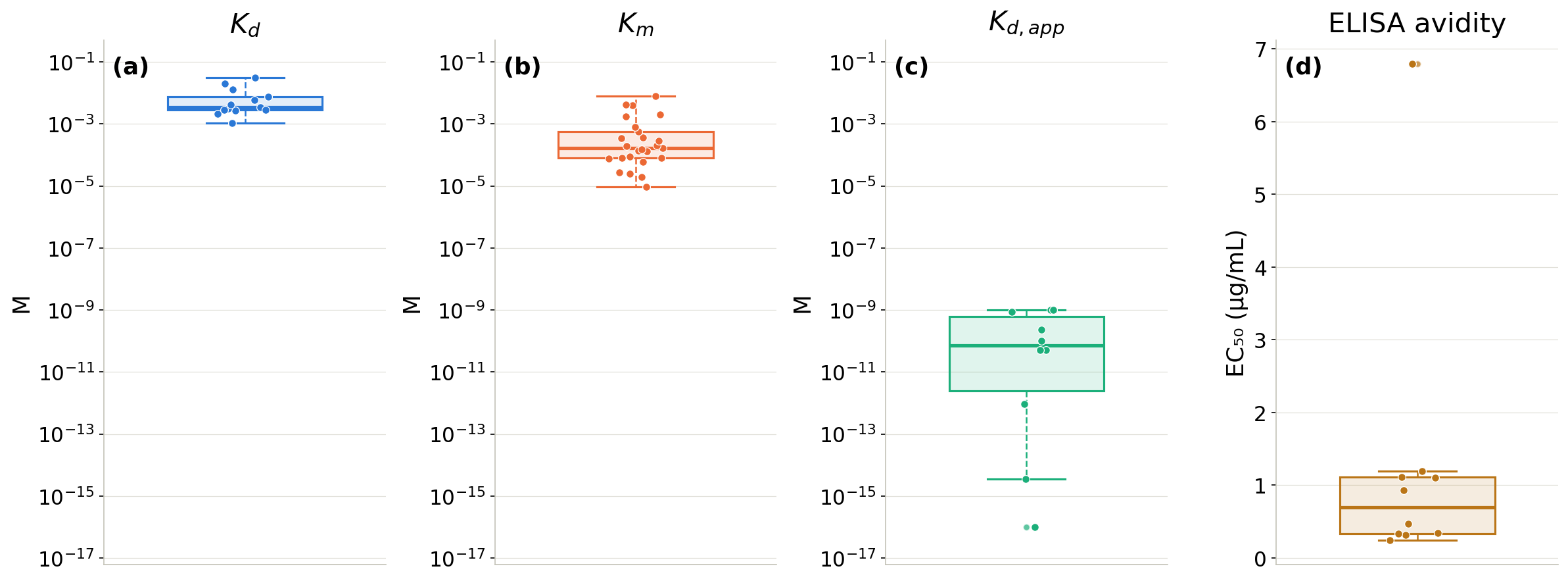}
    \caption{(a) NA-SIA dissociation constant \cite{sauter_hemagglutinins_1989, hanson1992proton, sauter1992crystallographic, xiong2013receptor, gamblin2021hemagglutinin, benton_biophysical_2015, reiter-scherer_force_2019}, (b) Michaelis constant \cite{cabezas1980neuraminidase, arora1986characterization, benton_biophysical_2015, benton2017role, xu_functional_2012}, (c) particle-surface dissociation constant \cite{benton_biophysical_2015, gamblin2021hemagglutinin}, and 
    (d) ELISA avidity \cite{xu_functional_2012}, as reported in experimental studies.}
    \label{Fig:database}
\end{figure}

Influenza virus–receptor binding kinetics have been investigated using surface plasmon resonance (SPR) and biolayer interferometry (BLI), which enable real-time measurements of association and dissociation at receptor-functionalized surfaces \cite{takemoto_surface_1996,benton_biophysical_2015,de_vries_influenza_2020,guo2018kinetic}. TIRF-based single-particle tracking approaches have further been used to resolve the attachment, mobility, residence time, and detachment of individual influenza virions on receptor-containing membranes \cite{muller_mobility-based_2019}. At the molecular level, single-molecule force spectroscopy, typically based on atomic force microscopy, and NMR have provided kinetic and mechanical information on individual and multivalent interactions between hemagglutinin or neuraminidase and sialic acid receptors \cite{sauter_hemagglutinins_1989,reiter-scherer_force_2019}. Together, these experimental approaches probe influenza virus binding across different spatial and temporal scales, from ensemble association and dissociation to the dynamics and rupture of individual virus–receptor contacts. 
Single-molecule spectroscopy and NMR studies allow for the estimation of the HA-SIA dissociation constant $K_d$ and the Michaelis constant $K_m$ (Fig.~\ref{Fig:database}a and b). Instead, experiments tracking whole-virion dynamics on conjugated surfaces measure the apparent dissociation constant ($K_d$) or avidity as defined in ELISA experiments (Fig.~\ref{Fig:database}c and d). The apparent dissociation constant is an emergent property that depends on system parameters such as the size of the virions, NA activity, receptor densities, {\em etc}. All of these parameters contribute to the variance of $K_{d,{\it app}}$ in Fig.~\ref{Fig:database}c, which is larger than that of $K_d$. Multivalent models (Sec.~\ref{Sec:Multivalent}) can link $K_d$ to $K_{d,{\it app}}$. In experiments in which NA is not inhibited, such models must consider the effect of NA activity (Sec.~\ref{Sec:Motility}). The dissociation constant of binding molecules tethered to supporting planes (cell or virion surface) is affected by configurational terms (Fig.~\ref{Fig:Bond Interaction}b Sec.~\ref{Sec:configurational}).

Estimations of the {\it on}/{\it off} reaction rates ($k^0_{\it on}$, $k^0_{\it off}$, $k^0_1$, and $k^0_{-1}$) tend to scatter depending on the study. Understanding of the magnitude of the reaction constants is important when modeling whole particle dynamics, which are also controlled by other timescales, such as the diffusion constant of the receptors.

\subsection{Secondary HA and NA binding sites}\label{Sec:second_HA_site}

{\bf HA:} A second binding site for SIA has been reported for both spike proteins. Using X-ray crystallography, Sauter {\em et al.\ }identified a second binding site for SIA in the lower part of the HA1 head domain of HA from A/X31 \cite{sauter1992crystallographic}. This site has an affinity for SIA that is at least four times weaker than that of the primary SIA binding site. Results obtained by Bandlow {\em et al.\ }using microscale thermophoresis support the existence of such a secondary binding site in strain A/X31 \cite{bandlow2017spatial}. Studies on the binding of DNA strands with a multivalent display of sialyl-LacNAc at defined intervals revealed a dissociation constant for the secondary binding site that was two orders of magnitude larger than that of the primary binding site (Fig.~\ref{Fig:database}a). Given this weak affinity, detecting the secondary binding site using monovalent SIA compounds is challenging. Bandlow {\em et al.\ }\cite{bandlow2017spatial} has done this through the binding of a multivalent SIA structure to the primary site, facilitating binding to the nearby secondary site. The relevance of the secondary binding site to infection has not yet been demonstrated. There is also no evidence as to whether such a binding site also exists for HA of other IAVs. The sequence of this site, as identified for A/X31, is not well conserved \cite{stray2000influenza}.

{\bf NA:} In contrast, a secondary SIA binding site on the neuraminidase of avian IAV has been identified \cite{uhlendorff2009functional,du20192nd,varghese1997structural,sun2014structure,lai2012secondary,sung_role_2010} and its relevance to infection has been proven \cite{du20192nd,du2021second}. This secondary binding site has been demonstrated for subtypes N1, N2, N5, N6 and N9. It is located in close proximity to the NA catalytic active site \cite{vavricka2013influenza}. Both (2,3)-linked SIA and (2,6)-linked SIA bind to the secondary binding site, with specificity or preference depending both on the subtype and on the method used \cite{uhlendorff2009functional,du2018substrate,dai2017mutation}. The binding affinity of the secondary site, at 5 - 20 mM ($K_d$), is significantly weaker than that of the catalytic site, at 0.1 - 1.0 mM ($K_m$) \cite{du2021second,mcauley2019influenza}. 
This extremely weak affinity is unlikely to be biologically relevant for monovalent SIA binding. 
However, in the case of multivalent binding, e.g. to IAV, the binding of SIA residues to the secondary site is facilitated and promote the catalytic activity of NA due to their proximity to the catalytic site \cite{du2021second}. The specific mechanism underlying the increase in NA catalytic activity mediated by the secondary site has not yet been fully elucidated. Nevertheless, the biological relevance of this increase in activity is demonstrated by mutations in the secondary binding site that lead to the inhibition of SIA residue binding and reduce virus replication (\cite{du2021second} and references therein).

\subsection{HA-NA balance}\label{Sec:HA-NA_synergy} 

As mentioned above, the secondary binding site of NA is highly conserved across all avian IAV NA subtypes. In contrast, this site is invariably absent in human IAVs and, consequently, in pandemic IAVs as well. Various studies have investigated different combinations of human and avian HA and NA with regard to their binding behaviour. These studies concluded that an optimal balance between the SIA-binding affinity of HA and the catalytic activity of NA (including its SIA-binding affinity) is required for viral replication. The interplay between HA and NA is also evident in substitutions in the HA secondary binding site that influence receptor-binding behaviour (\cite{du2021second} and references therein).

The reciprocal adaptation of HA and NA to maintain an optimal balance is not limited solely to the presence or absence of the secondary binding site, but has also been observed independently of this site in several studies. For example, Ilyushina {\em et al.\ }\cite{ilyushina2012decreased} investigated changes in HA and NA associated with the adaptation of H5N1 viruses to humans. Under the influence of NA inhibitors, mutations developed in HA and NA that reduced binding to (2,3)-linked SIA receptors, as well as reducing NA activity. This study was the first to demonstrate that reducing NA activity through NA inhibitors is essential for a pandemic H5N1 IAV to adapt to humans. The significance of this functional balance was also evident when NA inhibitor-resistant mutants were characterised. Unexpectedly, these resistant viruses did not develop mutations in the NA, but rather in the HA which resulted in reduced receptor affinity \cite{mcauley2019influenza,mckimm1996generation,mckimm1996mutation,mckimm1998mutations,mckimm-breschkin_resistance_2000}. Consequently, lower HA activity was required for efficient virion release. However, in the absence of NA inhibitors, the receptor affinity of some of these mutants was so low that the catalytic activity of NA was able to cleave the receptors before the viruses could form a stable bond with them via HA, allowing the viruses to be taken up into the cell. Another adaptation strategy to preserve the balance is to shorten the NA stalk, thereby reducing its activity \cite{blumenkrantz2013short,hermann2025clade,giannecchini2006comparison}. For more examples of keeping the balance see \cite{du2021second,mcauley2019influenza}.

Xu {\em et al.\ }proved the existence of a correlation between HA and NA activity in human IAV systems \cite{xu_functional_2012}. They plotted HA dissociation constants $K_d$ {\em vs.\ }NA catalytic rates $k_{\it cat}$ (Eq.~\ref{Eq:MM})
and showed how four pandemic strains can be fitted by a straight line. This work did not find a similar correlation in swine IAV systems \cite{xu_functional_2012}. A more recent study \cite{marougka2026} suggested that the balance between HA and NA in H1N1 prepandemic swine viruses adjusted before adapting to a human host. Such a change could be due to differences in the hosts' sialoglycome. Another explanation could be  that the evolutionary pressures in swine do not require a tight match between HA and NA activities. This could explain why swine can be infected by both human and avian IAV.

The key finding of these studies is that adaptations to the host cell occur alongside changes in the receptor affinity and catalytic activity of HA and NA, respectively. In other words, the activity of the two proteins must be balanced to enable optimal infection and efficient release from host cells. Otherwise, excessively high NA catalytic activity would cause the viruses to detach from the cell surface too early, thereby preventing them from entering the cell. Conversely, insufficient NA activity would, on the one hand, significantly slow down or even prevent the viral mobility on the cell surface and, on the other hand, to interfere with the release of newly formed virions from the cell surface. 
Consequently, in order to model the binding behaviour of IAV to the host cell and to multivalent inhibitors in a way that reflects reality, it is necessary to draw on experimental data for HA and NA receptor affinity, as well as for catalytic NA activity, which have been measured on the same viral strain. Sec.~\ref {Sec:Motility} presents biophysical models of virion motility and detachment. These models are parameterized by single bond reaction rates and affinities ($K_d$ and $K_m$, Fig.~\ref{Fig:database}). However, as discussed in the next section, the dissociation rates appearing in multivalent models are modulated by configurational terms.

\subsection{Configurational terms}\label{Sec:configurational}

Reaction kinetics between membrane- or scaffold-bound ligands and receptors differ fundamentally from those measured for freely diffusing molecules in solution. In the case of influenza virus, both HA and sialylated receptors are constrained by their respective surfaces, and binding therefore occurs within a reduced configurational space (Fig.~\ref{Fig:Bond Interaction}b).  Specifically, 2D single-bond equilibrium constants and rates are calculated by adding configurational terms to their corresponding 3D values 
\cite{varilly2012general,wu_transforming_2011,de_gernier_new_2014,xie_linking_2014,weikl_binding_2016,mognetti_programmable_2019}. Such configurational terms quantify the reduction in the number of possible configurations of the polymeric backbone following pairing of the supported SIA residue with an HA ligand \cite{varilly2012general,de_gernier_new_2014}.
Configurational terms are linked to the force exerted by a bond on the virion particle and depend on the length of the polymeric scaffold (supporting SIA, Fig.~\ref{Fig:Bond Interaction}b), which is variable \cite{peng_recent_2017,wang_structural_2019}. In addition, fluctuations of the opposing membrane surfaces can influence the probability of bond formation and dissociation \cite{xie_linking_2014,steinkuhler_membrane_2018}. Simulations and theoretical models are crucial in estimating these contributions, with the goal of quantifying single-bond reaction rates of supported ligands/receptors. Such rates are then used, at longer lengths and timescales, to describe the emerging dynamics of entire virions following the formation/rupture of many bonds (see following sections).

\section{Multivalent Models}\label{Sec:Multivalent}

Multivalent models are coarse-grained approaches that allow one to derive the dynamics of ligand-presenting particles starting from single-bond reaction dynamics (Sec.~\ref{Sec:LR}). Single-bond dynamics are controlled by configurational terms, which depends on the molecular architecture and relative distance between conjugated molecules (Sec.~\ref{Sec:LR} and Fig.~\ref{Fig:Bond Interaction}b). Instead, multivalent models quantify the emerging particle dynamics arising from many reaction events (either formation or rupture of a bond). Multivalent models are based on combinatorial calculations and naturally lead to the estimation, for instance, of the avidity in bioadhesion \cite{bell_models_1978,martinez-veracoechea_designing_2011,varilly2012general}.

Multivalency is central to several key steps of the influenza A virus infection cycle, including attachment to the host-cell membrane, lateral motion across receptor-containing surfaces, and the release of newly formed virions from infected cells. Multivalent interaction models are therefore useful for describing how the collective formation and rupture of many weak HA-SIA contacts gives rise to stable adhesion and dynamic virus motion.

The same principles are exploited in the design of antiviral nanomaterials, including functionalized polymers, nanoparticles, and dendrimers. By presenting multiple virus-binding ligands, these materials can engage several viral surface proteins simultaneously and thereby achieve high-avidity binding. Depending on their architecture, they can competitively block receptor binding, sterically shield the virion surface, or promote virus aggregation, following reducing productive interactions with host-cell receptors (see Sec.~\ref{Sec:Nanom}).

\subsection{Theoretical and Simulation Models}\label{Sec:Methods}

Modeling the dynamics of an entire virion at atomistic resolution over biologically relevant length and time scales is to date computationally impossible. Instead, coarse-grained approaches (Fig.~\ref{Fig:General_Model}c) map ligand-receptor complexes into simplified bonds which are modeled, for instance, by means of harmonic springs. Such bonds are not static but are reversibly added and removed from the system at given rates ($k_{\it on}$, $k_{\it off}$, $k_1$, and $k_{-1}$ in Fig.~\ref{Fig:Bond Interaction}b), as obtained from the corresponding bulk values ($k^0_{\it on}$, $k^0_{\it off}$, $k^0_1$, and $k^0_{-1}$ in Sec.\ \ref{Sec:LR}) by considering configurational terms (Sec.\ \ref{Sec:configurational} and Fig.~\ref{Fig:Bond Interaction}b).

Equilibrium theories of multivalency \cite{bell_models_1978,martinez-veracoechea_designing_2011,varilly2012general} use statistical mechanics to describe the collective adhesion arising from many reversible ligand-receptor interactions. They are typically parametrized by the dissociation constant $K_d$ of a single ligand-receptor pair (see Sec.\ \ref{Sec:LR}) along with other biophysical parameters (for instance, ligand and receptor densities).

Molecular simulations of coarse-grained models featuring reversible bonds are challenging because of the need to dynamically change the force field to account for the formation/rupture of ligand-receptor bonds. Reaction-diffusion methods combine molecular dynamics to update the system configuration at a given set of bonds with algorithms to implement ligand/receptor reactions \cite{martinez-veracoechea_designing_2014,kowalewski_multivalent_2021,lowensohn_sliding_2022,marbach_nanocaterpillars_2022,stevens_sliding_2023}. To the best of our knowledge, a systematic comparison between different reaction-diffusion methods that assesses the tradeoff between their limitations and computational cost is still missing.

Reaction-diffusion simulations allow studying kinetic aspects of multivalent interactions, which are not captured by equilibrium theories. In general, multivalent theories are equilibrium models parametrized by the dissociation constant ($k_{\it off}/k_{\it on}$, see Fig.~\ref{Fig:Bond Interaction}b). They do not depend on the absolute values of the reaction rates. In particular, they assume that the binding/unbinding events are much faster than any other timescale entering into the system. As experimentally and theoretically studied by Lanfranco {\em et al.\ }\cite{lanfranco_kinetics_2019}, this approximation fails under strong binding conditions in systems where the ligands outnumber receptors. In particular, small values of $k_{\it off}$ result in steady states featuring fewer surface-bound particles than are predicted by equilibrium models. This is because early-bound particles tend to deplete the surface from receptors. The binding of extra particles (which would lower the free energy due to a combinatorial or avidity term) is kinetically limited by the lack of receptors, which are found clustered under the early-bound particles and cannot detach because of the small value of $k_{\it off}$. 
Related kinetic effects have been observed in the self-assembly of multivalent particles, where slow bond dissociation can limit structural rearrangement and thereby constrain aggregate size \cite{bachmann_bond_2016,ranganathan_dynamic_2020}.

Reaction diffusion simulations have been employed to study IAV particles dynamics mediated by reversible HA-SIA bonds \cite{vahey_influenza_2019,stevens_sliding_2023,agarwal_kinetics_2024,kolbe_understanding_2025}. These studies also account for the action of NA enzymes by implementing reactions that reduce the number of receptors within the reach of NA ligands. 
In existing studies, the Michaelis–Menten (Eq.\ \ref{Eq:MM}) reaction is simplified into a single-step reaction by taking the limit in which the catalytic rate is much larger than the {\it on} and {\it off} rates, $k_{\it cat} \gg k_1, k_{-1}$ (Fig.\ \ref{Fig:Bond Interaction}b)  
\begin{eqnarray}
\ce{NA + SIA ->[k_d \approx k_1] NA + SIA^*}
\qquad .
\end{eqnarray}
In the future, it will be important to consider the full Michaelis–Menten reaction (Eq.\ \ref{Eq:MM}) to assess the role of the intermediate complex (NA-SIA) in the adhesion term. This is particularly relevant because NA-mediated cleavage is not instantaneous. Catalytic turnover numbers reported for influenza NA are typically on the order of tens of s$^{-1}$, although they vary considerably with NA subtype, viral strain, substrate, and experimental conditions. The lifetime of the NA-SIA complex is determined by both catalytic cleavage ($k_{\it cat}$, see Eq.~\ref{Eq:MM}) and substrate dissociation ($k_{-1}$, Fig.\ \ref{Fig:Bond Interaction}b), and may therefore provide a finite time window during which NA contributes to virion-surface adhesion. Indeed, agglutination of IAV-like particles lacking HA has been reported \cite{hooper_mutant_2013}, supporting the concept that NA does not act exclusively as an antagonist of virus–receptor binding.
Currently, HA ligands are also mapped into point-like units, neglecting the trimeric nature of the receptor-ligand interaction. 
More realistically, several sialylated glycans may bind simultaneously to different receptor-binding sites within the same HA trimer, and sufficiently long glycans may even bridge two HA subunits \cite{peng_recent_2017}. In addition to this structural simplification, reaction–diffusion models often treat binding and unbinding events as Poisson processes when ligands are represented implicitly.
This is an approximation that neglects rebinding events, i.e., the fact that a ligand-receptor complex is more likely to reform immediately after unbinding with the two interaction partners still in proximity of each other \cite{van_zon_greens-function_2005}.

\begin{figure}
    \centering
    \includegraphics[angle=0,width=0.95\linewidth]{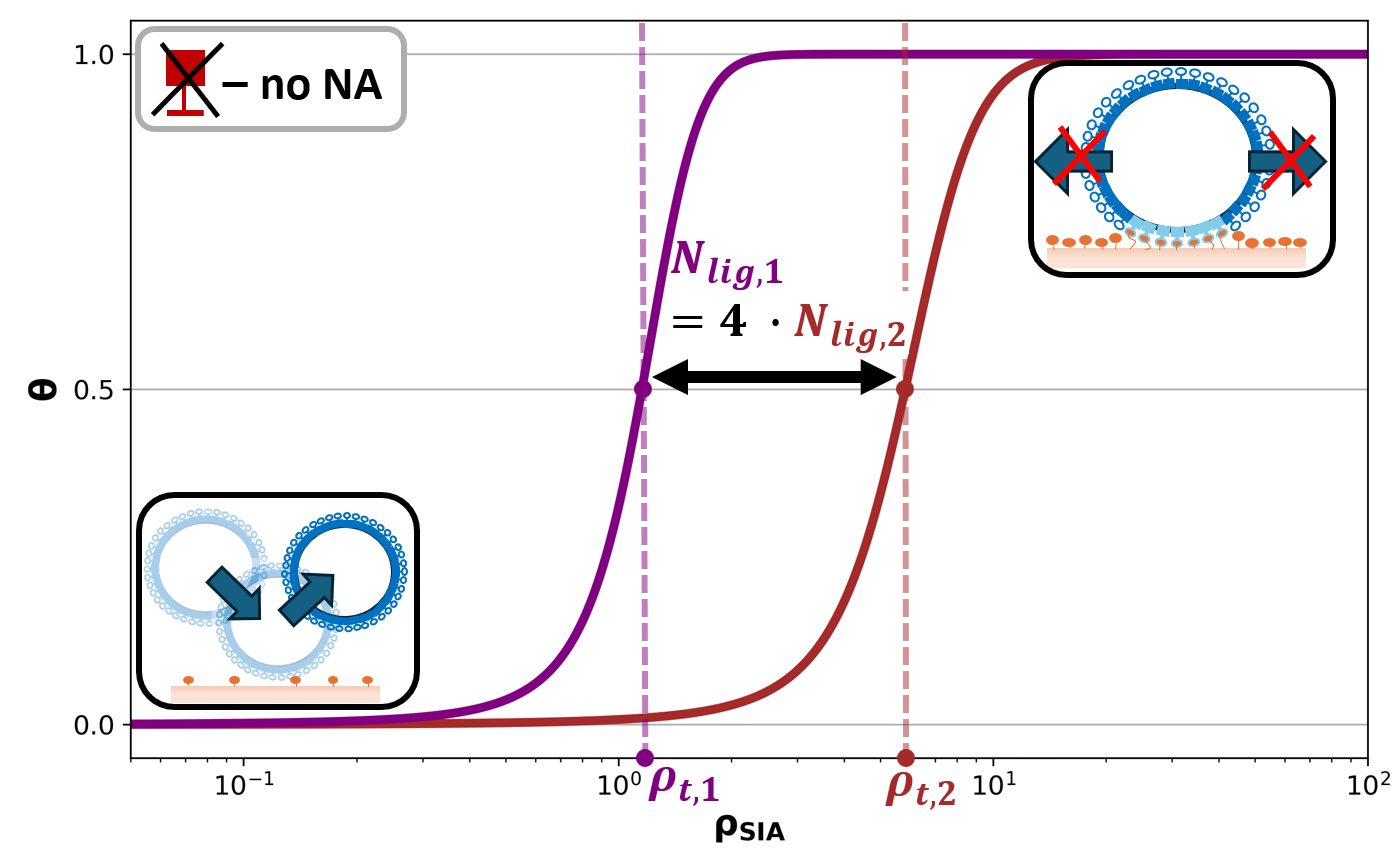}
    \caption{Schematic representation of the coverage of a receptor-presenting surface binding multivalent particles carrying conjugated ligands. For multivalent particles, the transition between uncovered ($\theta=0$) and fully covered surfaces ($\theta=1$) develops within narrow intervals of receptor concentrations ($\rho_{\it SIA}$). The transition density is a function of the number of ligands ($N_{\it lig}$). When increasing the number of ligands (in the figure by a factor 4), the transition density $\rho_t$ decreases \cite{martinez-veracoechea_designing_2011}. In a pleomorphic system, the number of ligands follows a broad distribution. Accordingly, at a given surface concentration, most of the particles are either stuck at the surface or never bind the surface. This scenario is detrimental to infection spreading, as it limits trafficking from and towards the cell surface. When bound ($\theta=1$), the particles cannot laterally diffuse, unless close to the transition point \cite{marbach_nanocaterpillars_2022,lowensohn_sliding_2022} or anchored to mobile receptors.  }
    \label{Fig:Superselectivity}
\end{figure}

Reaction kinetics have an important effect on the motility of the virion particles. Multivalent interactions are sticky: A particle binding to a surface presenting receptors with fixed anchoring points can hardly move without detaching, except when fine-tuning the system's parameters close to the detachment point \cite{lowensohn_sliding_2022,marbach_nanocaterpillars_2022} (Fig.~\ref{Fig:Superselectivity}).  Motility of the particle at the cell membrane arises either from the motility of the receptors or from the effect of catalytic activity, as described in Sec.\ \ref{Sec:Motility}.

\subsection{Adhesion to Functionalized Surfaces: Superselectivity}\label{Sec:RB}

At the single-particle scale, multivalent models (Sec.~\ref{Sec:Methods}) are used to quantify the adhesion strength of virions (or, in general, ligand-presenting particles) to the cell surface. 
%Adsorption profiles (e.g., calculated using Langmuir isotherms) are inferred from the particle-surface and particle-polymer adhesion free energy and the dissociation constant $K_d$, the bonding potential (e.g., harmonic), and the positions of the receptors and ligands on the virion and cell surface.Many works have studied targeting of the cell surface by functionalized particles. 
The adhesion free energy is affected by relevant biophysical parameters, including the concentrations of ligands and receptors \cite{martinez-veracoechea_designing_2011}, the presence of non-specific interactions between ligands/receptors \cite{liu2020combinatorial,tian_design_2020,di_michele_steric_2018}, multiple types of ligands \cite{curk_optimal_2017,angioletti-uberti_exploiting_2017}, ligands that form multimeric complexes \cite{di_michele_communication_2016} (possibly due to linker molecules that cross-link ligands with receptors \cite{xia_designing_2024}), and the particle's deformability \cite{hamzeh_single_2025}.

In general, multivalent particles featuring weak bonds can bind functionalized surfaces {\em superselectively} \cite{martinez-veracoechea_designing_2011,dubacheva_designing_2015}. Specifically, they typically respond sharply to variations in receptor concentrations (Fig.~\ref{Fig:Superselectivity}). In the simple binding mode of monovalent ligands, the fraction of bound ligands increases approximately linearly with ligand density. In contrast, multivalent binding is characterized by cooperative effects, such that binding increases more than linearly with ligand density. This cooperativity can give rise to a steep, almost step-like transition from a state in which essentially no receptors are bound to one in which nearly all particles are bound, thereby resulting in superselective binding. This result enables the design of multivalent probes selectively targeting surfaces that express a receptor concentration greater than a specific threshold. More advanced designs allow targeting a finite range in receptor concentrations \cite{tian_design_2020} or surfaces that simultaneously express multiple types of ligands \cite{curk_optimal_2017}.
SIA-HA bonds are weak. The interaction between virions and the cell surface is cooperative and therefore in regimes where superselectivity is at play \cite{hamming2023receptor}.

Superselectivity is a powerful concept that is helping to design delivery vectors. Superselectivity is also at play in biological systems (e.g., \cite{young_triad_2025,schmidt2015liquid}) and likely affects the trafficking dynamics of virions across the cell surface. However, as highlighted in Secs.\ \ref{Sec:Lim_Multi} and \ref{Sec:Motility}, the targeting properties of IAV virions cannot be entirely understood without considering the activity of the NA. Some hints in this direction are given in Fig.~\ref{Fig:Superselectivity}. In superselectivity, the threshold concentration is a function of the particle size (or the number of ligands on the particle). In a pleomorphic system like IAV, most of the virion particles would be either stuck at the cell surface (without the possibility of detaching or moving, Sec.~\ref{Sec:Motility}) or not bound at all. This scenario is not optimal for infection propagation, which is expected to be sustained by trafficking of virions from infected towards healthy cells. In Sec.~\ref{Sec:Motility}, we review how NA activity can alter the scenario of Fig.~\ref{Fig:Superselectivity} and favor trafficking.

\begin{figure}
    \centering
    \includegraphics[angle=0,width=0.99\linewidth]{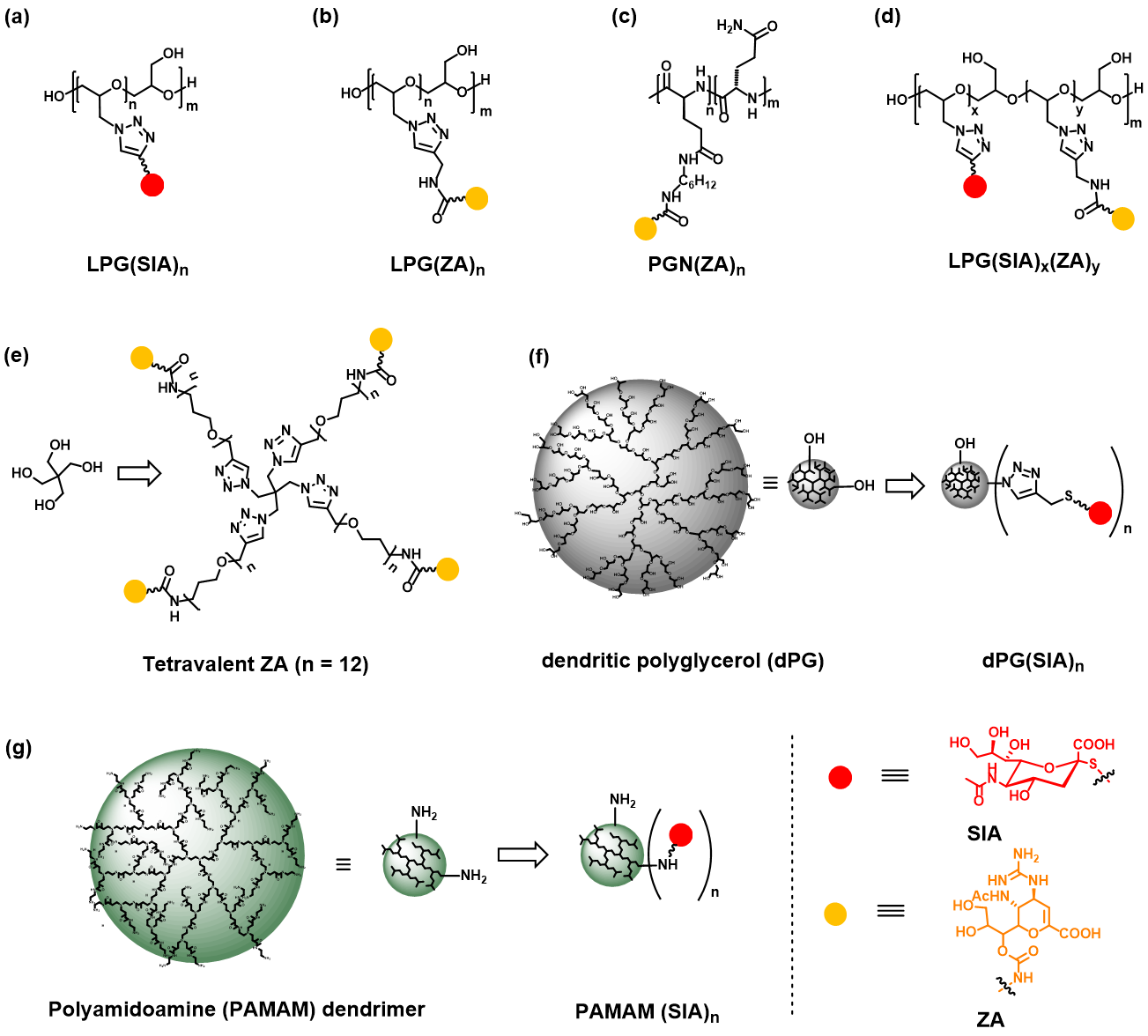}
    \caption{Representative multivalent glycoside architectures developed as inhibitors of IAV. Chemical structures of (a) linear polyglycerol functionalized with sialic acid, LPG(SIA)$_n$ \cite{bhatia_linear_2017}; (b) zanamivir-functionalized linear polyglycerol, LPG(ZA)$_n$ \cite{parshad2023dual}; (c) poly-L-glutamine-based polyvalent zanamivir, PGN(ZA)$_n$ \cite{wang2011polymer,lee2012polymer}; (d) heteromultivalent linear polymer co-functionalized with sialic acid and zanamivir, LPG(SIA)$_x$(ZA)$_y$ \cite{parshad2023dual}; (e) structurally defined tetravalent zanamivir containing four ZA units connected through PEG-based spacers \cite{fu2016structure}; (f) dendritic polyglycerol functionalized with sialic acid, dPG(SIA)$_n$ \cite{bhatia_linear_2017}; and (g) sialylated polyamidoamine (PAMAM) dendrimer, PAMAM(SIA)$_n$ \cite{kwon2017nanostructured}. Red and yellow symbols denote SIA- and ZA-containing functionalities, respectively.}
    \label{Fig:Antiv}
\end{figure}

\section{Designing Multivalent Glycosides as IAV Inhibitors}\label{Sec:Nanom} 

Different polymers and nanoparticles decorated with SIA derivatives have been designed and studied for their ability to compete for virus binding to host cells. Some representative structures of antiviral multivalent glycosides are provided in Fig.~\ref{Fig:Antiv}.
As compared to the ideal superselective trend (Fig.~\ref{Fig:Superselectivity}), the inhibitory activity of these constructs does not vary monotonically with the number of SIA units \cite{bhatia_linear_2017}. Quantitative understandings of the organization of IAV virus-conjugated receptors at the cell surface have inspired inhibitor designs.
While a typical spherical IAV particle with a diameter of 120$\,$nm bears around 300–400 HA trimers and 30-50 NA tetramers on its surface, there are around 50-200 SIA per 100$\,$nm$^2$ on the host cell surface. In general, the design of multivalent inhibitors involves several key considerations that determine their ability to engage viral receptors effectively. Multivalent inhibitors should mimic the natural multivalency of the virus–host cell interaction by presenting a large number of monovalent ligands units on a suitable scaffold, while carefully tuning their spatial density and accessibility. The scaffold should be sufficiently flexible to allow individual ligands to reach and engage multiple neighbouring receptors, thereby enabling multivalent interactions and increasing the overall binding energy. However, excessive flexibility can result in a substantial entropic penalty upon binding due to the loss of conformational freedom, thereby weakening the overall interaction or even cause repulsive interaction~\cite{krishnamurthy2006, dubacheva2014,liu2020combinatorial}. Similarly, excessively high ligand densities may lead to steric hindrance between neighbouring ligands and reduce receptor accessibility~\cite{Hlavacek1999,shi2007}. In addition to these binding effects, the physical size of the inhibitor and its scaffold can generate steric shielding and cover regions of the virus surface beyond the area directly contacted by the inhibitor, thereby further reducing binding to the host cell~\cite{vonnemann2015,bhatia_linear_2017}. The interplay between multivalent binding, conformational entropy, ligand accessibility, and steric shielding ultimately determines the optimal scaffold architecture and SIA density for selective IAV binding.

The role of optimal SIA density on two polymers with identical chemical composition but distinct architectures—linear and dendritic—was highlighted in the study by Bhatia et {\em al.}~\cite{bhatia_linear_2017}. Notably, the optimal SIA densities for both polymer types were approximately 30–70 SIA per 100$\,$nm$^2$, which falls within the same range observed on host cell surfaces.  The optimized multivalent sialosides demonstrated up to 15,000-fold enhanced activity compared to the monovalent $\alpha$-2,6-sialyllactose in the hemagglutination inhibition (HAI) assay. At the same time, for linear and dendritic multivalent sialosides with comparable binding affinity to the virus, inhibition potential of linear compounds was up to 16 times higher than the dendritic analogs. This behaviour can be attributed to the greater steric shielding provided by a more flexible linear scaffold, which aligns with the phenomenon first described by Sigal et~{\em al} \cite{sigal1996polyacrylamides}, where a linear random coil architecture sterically shields unbound receptors while enabling efficient multivalent ligand–receptor interactions.
Numerous multivalent sialosides based on polymers \cite{mammen1995effective,kwon2017nanostructured}, nanoparticles \cite{lauster2020phage,bhatia2020adaptive}, and proteins \cite{wang2013design} have been developed to overcome the low binding affinity of monovalent SIA for HA through multivalent effect. 
Nevertheless, most polysialosides still lack broad activity and high efficacy largely because of considerable structural and amino acid sequence diversity of the HA binding pockets of different IAV strains. Although replacing SIA with more natural 6$'$-sialyllactose (6$'$-SL) broaden the antiviral activity \cite{stadtmueller2021evaluation}, achieving high efficacy remains a significant challenge.

In addition to statistical multivalent displays, structurally defined scaffolds have been developed to control the number, spacing, and spatial arrangement of sialoside ligands. Bandlow {\em et al.\ }\cite{bandlow2017spatial} used self-assembled PNA–DNA scaffolds to systematically vary the distance between two sialyl-LacNAc ligands and thereby probe the spatial requirements for multivalent HA binding. In contrast to highly flexible PEG linkers, the more rigid PNA–DNA constructs revealed a pronounced distance dependence, with optimized bivalent arrangements producing binding enhancements of up to three orders of magnitude compared with the monovalent ligand \cite{bandlow2017spatial}. This concept was subsequently extended by assembling optimized PNA-DNA binders on long DNA templates generated by rolling-circle amplification, enabling simultaneous engagement of multiple HA trimers on the virion surface \cite{bandlow2019sialyl}.

An alternative strategy uses rigid protein-based scaffolds to control the spatial presentation of receptor mimetics. Sialylated Q$\beta$ bacteriophage capsids were designed to present SIA ligands in arrangements matching the geometry of the receptor-binding sites within the HA trimer. The resulting particles showed potent inhibition of IAV entry, with cryo-electron tomography revealing extensive coverage of the viral surface by bound Q$\beta$ capsids, indicating that geometrically matched multivalent binding is complemented by steric shielding of the virion \cite{lauster2020phage}. In a subsequent study, partially sialylated Q$\beta$ capsids retained antiviral activity, and heterobifunctional capsids displaying two different sialic-acid ligand–linker architectures showed enhanced inhibition compared with the corresponding monofunctional constructs \cite{adam2022design}. Together, these studies demonstrate that the antiviral properties of multivalent scaffolds can be tuned not only through valency and ligand density, but also through spatial organization, linker architecture, and the simultaneous presentation of distinct ligand configurations.

Considering the conserved nature and crucial role of NAs for virus detachment, many have also focused on the development of polyvalent zanamivir (ZA) as anti IAV strategy to block the infection process, where ZA is a commercial anti IAV drug that inhibits the enzymatic activity of NA generally at low nM concentration in vitro.
% suggested references: [mckimm-breschkin_resistance_2000] (McKimm-Breschkin 2000, Antiviral Res. -- review of influenza resistance to neuraminidase inhibitors, relevant to NA inhibitor context)
For example, a study by Lee {\em et al.} \cite{lee2012polymer} demonstrated that multivalent zanamivir attached via flexible linker to poly L-glutamine (PGN) has been up to 10000-fold more potent than ZA alone in plaque reduction assays. In the experimental set-up where inhibitors were incubated at different time points, it was shown that polyvalent zanamivir was able to inhibit infection in both early and late stage of infection, showing their interference in both entry and release mechanisms. While the polyvalent zanamivir developed by Lee {\em et al.\ }bears zanamivir statistically distributed onto the polymer chain, a precisely designed tetravalent zanamivir was developed by Fu {\em et al.\ }\cite{fu2016structure} that exhibited a marked increase of NA binding and inhibiting its enzymatic activity, and high antiviral efficacy over zanamivir. Especially, the lead tetravalent compound where four ZA molecules are attached to the core by do-decavalent PEG spacer showed up to 10000 times lower dissociation constant ($K_d = 4.6\,$pM) for NA as compared to monovalent ZA ($K_d = 1.1\,$nM). The lower $K_d$ value for tetravalent ZA was attributed to a 3-fold decrease in $k_{\it off}$ rather than $k_{\it on}$ when measured by SPR \cite{fu2016structure}.

Beyond homomultivalent analogues targeting HA or NA, Badri {\em et al.}~\cite{parshad2023dual} demonstrated the advantages of heteromultivalency using a linear polymer scaffold. 
Co-displaying 6$'$-SL and ZA on the same polymer produced substantially greater antiviral activity than homomultivalent polymers bearing either ligand alone or their corresponding physical mixtures. Importantly, these heteromultivalent polymers showed broad-spectrum activity against multiple IAV strains, where homomultivalent analogs showed little to no effect. 
Moreover, the lead heteromultivalent construct exhibited approximately 10,000 times higher potency in low nM concentrations than commercial zanamivir in both {\em in vitro} and {\em ex vivo} models. 
In these heteromultivalent systems, inhibition of NA enzymatic activity remained comparable to that of ZA, while inhibition of cellular infection was enhanced by approximately four-fold relative to commercial ZA. These findings suggest that the superior efficacy of the heteromultivalent analogues arises primarily from increased polymer adsorption and steric blockade of the viral surface, rather than from ligand–receptor binding alone. In future work, it would be valuable to further refine the design of heteromultivalent ligands that are equally capable of effectively reducing NA enzymatic activity and sterically blocking the IAV surface.

On the modeling side, studying targeting by polymers is more challenging because of the need to control configurational terms due to the backbone flexibility. A few studies consider targeting of flat surfaces \cite{de_gernier_new_2014,ravnik_designing_2024,curk_computational_2020}. In the future, it will be important to extend these cases to targeting spherical particles, and to characterize the role played by the particle's curvature. Experimental studies proved the advantages of designing nanomaterials matching the topology of targeted IAV virions \cite{nie2021heteromultivalent,nie2020topology}.
Moreover, to our knowledge, a characterization of the steric repulsion generated by the adsorbed layer of polymers onto the surface of the particles is also missing. This understanding will help clarify the results described above \cite{bhatia_linear_2017}, disentangling the effect of SIA density on the amount of adsorbed polymer (similar to what done in \cite{liu2020combinatorial} for nanoparticles) and the polymer configuration optimizing steric repulsion.

\section{Dynamics of IAV Virions in the Extracellular Medium}\label{Sec:Lim_Multi}

The phenomenology presented in Sec.~\ref{Sec:Multivalent} and Sec.~\ref{Sec:Nanom} does not account for the catalytic activity of NA enzymes. With catalytic activity (and in the absence of receptor replenishment), the system never reaches a steady state because the environment is constantly depleted from receptors. As a consequence, the interaction of the virions with their environment is not static but evolves in time. % This has important consequences on the ability of the host particles to navigate the extracellular medium (see Sec.~\ref{Sec:Motility}).
Key steps in the propagation of infection, such as the ability of the virions to crawl and detach from cell membranes or to penetrate the mucus, are expected to be regulated by the activity of the NA as shown below.

\begin{figure}
    \centering
    \includegraphics[angle=0,width=0.95\linewidth]{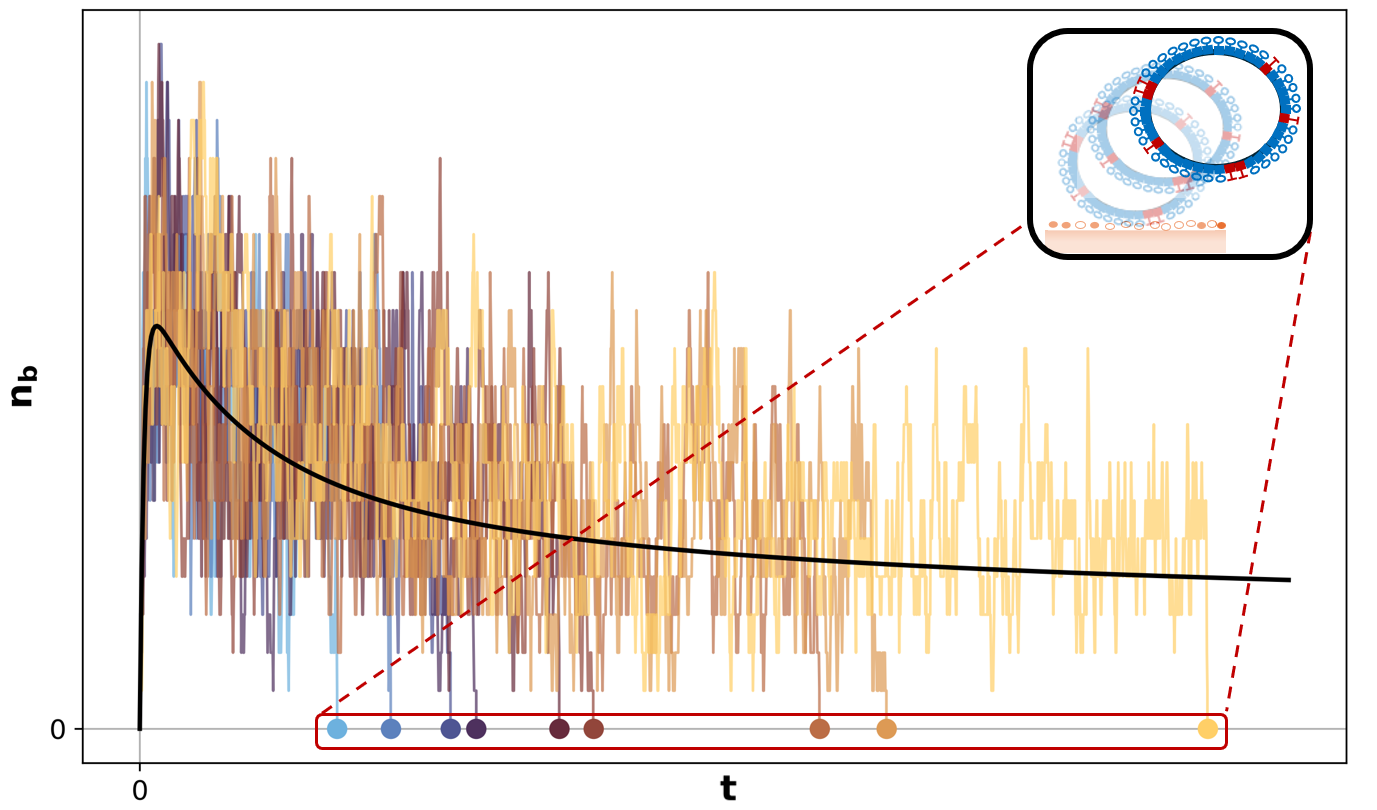}
    \caption{Upon encounter with the cell surface (at time $t=0$), the number of ligand-receptor bonds ($n_b$) readily increases before slowly decreasing due to NA activity. Thin, colored lines represent simulation trajectories (Sec.~\ref{Sec:Methods}). The particles detach when hitting the $n_b=0$ condition (dots). The thick black line represent a mean-field prediction for the number of bonds at a given receptor concentration (the latter evolving in time, Fig.~\ref{Fig:RhoSIA_t}). Results pertinent to surfaces with mobile receptors.}
    \label{Fig:Nb_t_Peak}
\end{figure}

\subsection{Virion Detachment from the Cell Surface}

M\"uller {\em et al.\ }\cite{muller_mobility-based_2019} used single-particle tracking to study the residence time of IAV virions bound to reconstructed cell surface models, constituted by supported-lipid bilayers functionalized with SIA residues. Among other results, this contribution reported how the residence time in the presence of catalytic activity is non-monotonic in the number of SIA-HA bonds, the latter inferred from the virions' lateral diffusion constant \cite{block_quantification_2016}. This result is surprising because, in general, multivalent theories predict that the adhesion strength is positively correlated with the number of ligand--receptor bonds \cite{bell_models_1978,varilly2012general}. Theoretical and numerical analysis have clarified this finding as due to a non-monotonic trend of the average number of HA-SIA bonds as a function of time (Fig.~\ref{Fig:Nb_t_Peak}) \cite{kolbe_understanding_2025}.  In particular, following attachment to the cell surface, the number of bonds increases rapidly before slowly decreasing due to the reduction in available receptors. In general, the residence time distribution is not Poissonian (like in systems without receptor-cleaving ligands) but features a peak whose position is controlled by the reaction rates (Sec.\ \ref{Sec:LR}) and the diffusion constant of the receptors \cite{kolbe_understanding_2025}. An optimal range of residence times is likely functional to infection spreading: If residence times are too long, virus particles cannot leave infected cells and spread the infection; if too short, particles may detach from the surface before triggering endocythosis and infect the cell. The results of Ref.\ \cite{kolbe_understanding_2025} may then provide a quantitative justification of the synergic activity between NA and HA.

\begin{figure}
    \centering
    \includegraphics[angle=0,width=0.95\linewidth]{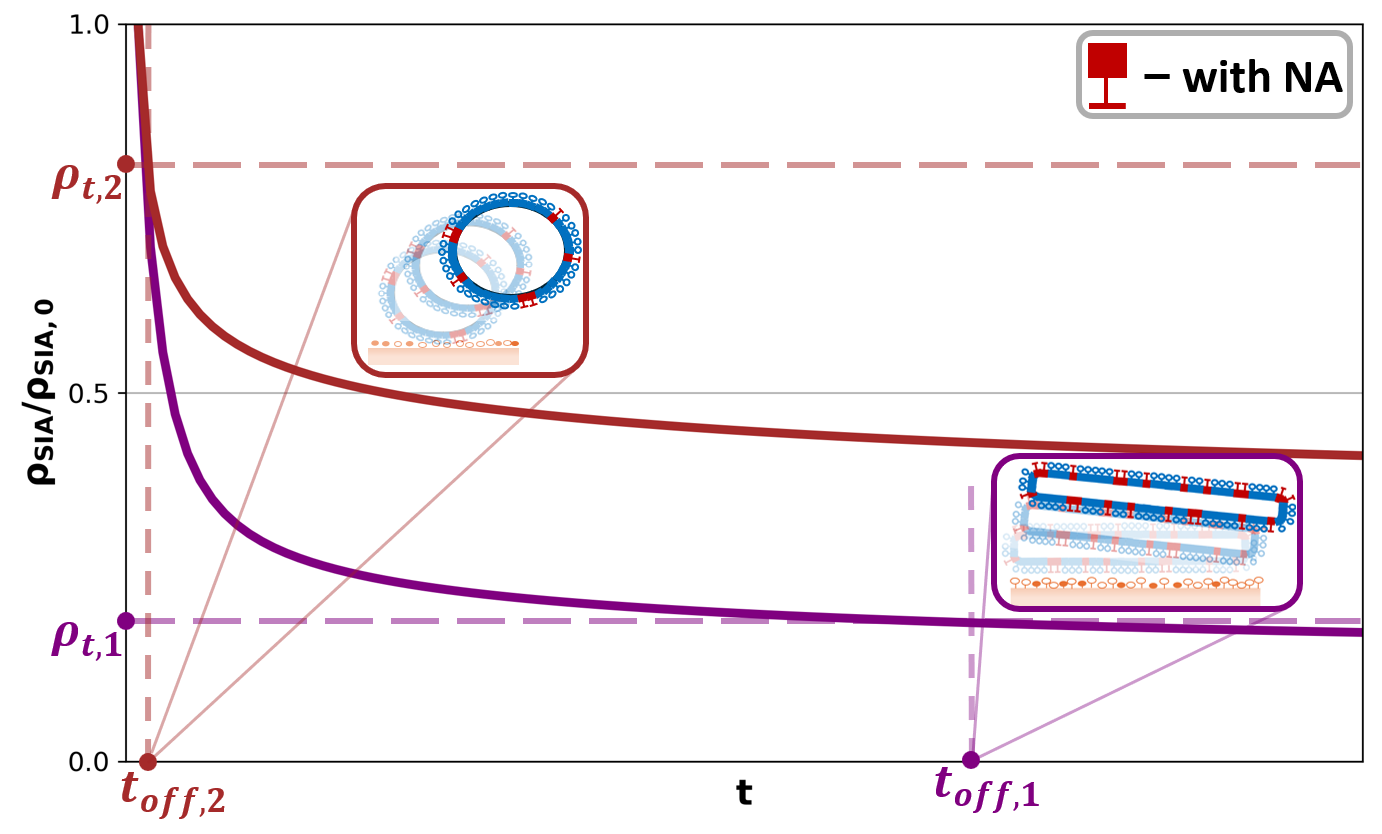}
    \caption{Time evolution of the SIA local density found under the particle as a function of the residence time $t$. $\rho_{\it SIA}$ decreases due to NA activity. Modulo stochastic fluctuations, particles detach from the surface at time $t_{\it off}$, when $\rho_{\it SIA}$ reaches the superselectivity threshold density $\rho_t$ (Fig.\ \ref{Fig:Superselectivity}). The bigger, filamentous particle depletes SIA more efficiently but detaches later than the spherical particle ($t_{\it off,1}>t_{\it off,2}$), at smaller values of $\rho_{\it SIA}$ ($\rho_{t,1}<\rho_{t,2}$). Results pertinent to surfaces with mobile receptors.} 
    \label{Fig:RhoSIA_t}
\end{figure}

As highlighted in Sec.\ \ref{Sec:RB}, multivalent particles are superselective \cite{martinez-veracoechea_designing_2011}. In general, superselectivity limits the dynamics of the system: except in fine-tuned conditions (for instance, in a narrow interval of receptor concentrations), most of the time, a particle will be found firmly bound to a surface or completely detached (Fig.~\ref{Fig:Superselectivity}). However, in IAV systems, because of the action of NA, the local receptor concentration sampled by a virion ($\rho_{\it SIA}$) steadily decreases as a function of time (Fig.~\ref{Fig:RhoSIA_t}). Virions of different sizes will then detach when reaching their corresponding threshold concentration $\rho_t$ (Fig.~\ref{Fig:RhoSIA_t}). Intriguingly, superselectivity is expected to reduce the variance of the detachment times. It will be relevant to test how this finding could impact multicellular models of infection (e.g., \cite{jacobs2019incomplete}).

\begin{figure}
    \centering
    \includegraphics[angle=0,width=\linewidth]{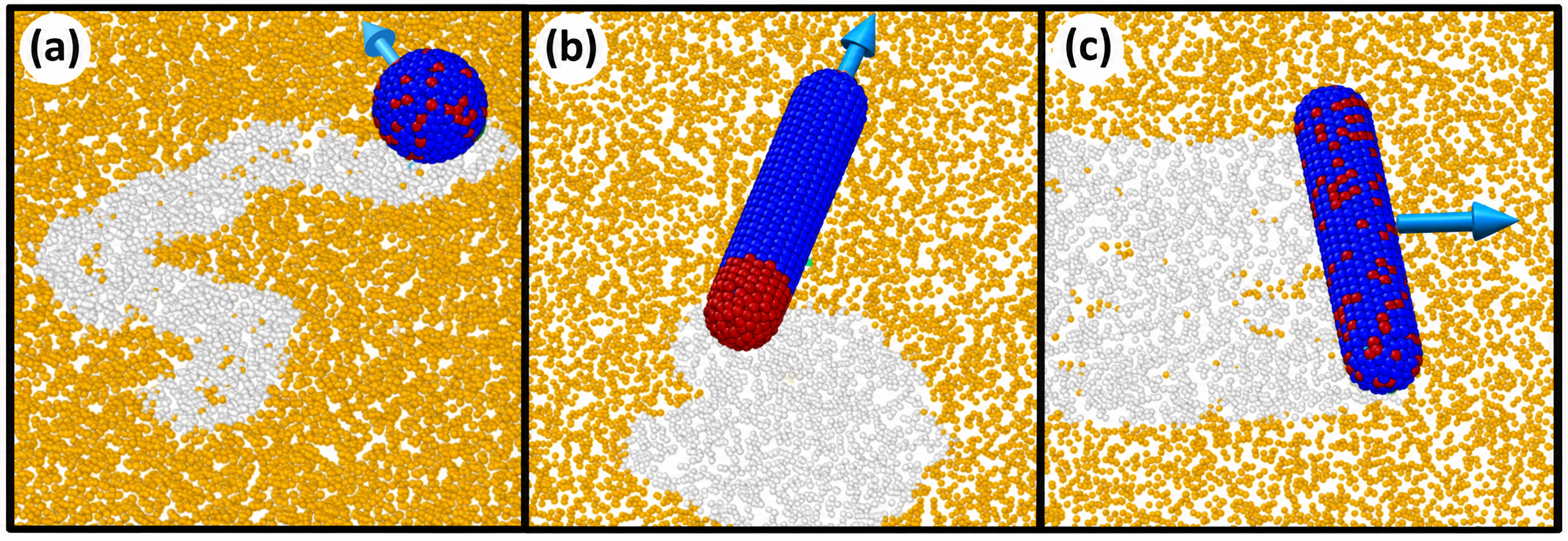}
    \caption{Influenza A Virion motility modes at the cell surface with fixed receptors. (a) In general, particles express directional motion away from their desialylated trail (gray circles) with random changes and drifts in direction. For filamentous virions \cite{vahey_influenza_2019,stevens_sliding_2023}, the spatial organisation of NA molecules at the virion surface determines the direction of motion. (b)
    Polarized distributions induce motion away from the NA-rich pole whereas
    homogeneous distributions (c) drives motion perpendicular to the particle's long axis. Created with Ovito \cite{ovito}.
    }\label{Fig:Motility}
\end{figure}

\subsection{Crawling of IAV Virions at the Cell Surface}\label{Sec:Motility}

Beyond facilitating detachment from the cell surface, the catalytic activity of NA is believed to enhance the motility of the virions. This is now established for virions moving across surfaces decorated with receptors with fixed tethering points \cite{sakai_influenza_2017,vahey_influenza_2019,agarwal_kinetics_2024,stevens_sliding_2023}. Instead, the effect of NA on the motility of the virions in the mucus is still debated (Sec.\ \ref{Sec:Mucus}) \cite{cohen_influenza_2013,agarwal_kinetics_2024,iseli_neuraminidase_2023,kaler_influenza_2022,kaler_mucus_2025}.

Multivalent particles carrying only receptor-binding ligands cannot diffuse across surfaces (with fixed ligands) while remaining bound. This is quantified by extremely small lateral diffusion constants \cite{lowensohn_sliding_2022,marbach_nanocaterpillars_2022} which has been estimated to scale like 
\begin{eqnarray}
D \sim k_{\it off} \lambda^2 / (n_b)^\alpha
\qquad 
\alpha=1,\quad 2
\label{Eq:Diffusion}
\end{eqnarray}
where $k_{\it off}$ is the rate at which ligand-receptor bonds are broken (Fig.\ \ref{Fig:Bond Interaction}b), $\lambda$ a microscopic length scale comparable to the bond length, $n_b$ the number of bonds, and $\alpha$ a parameter that likely depends on microscopic details of the model, which has been estimated in two extreme cases to be equal to $\alpha=1$ \cite{marbach_nanocaterpillars_2022} and $\alpha=2$ \cite{lowensohn_sliding_2022}. Using $\lambda \approx 1-10\,$nm, $k_{\it off}\approx 1\,$s$^{-1}$, and $n_b\approx 10$, Eq.\ \ref{Eq:Diffusion} predicts that it can take hours for a 100$\,$nm diameter particle to diffuse a distance comparable with its own size. This observation quantifies the notorious stickiness of the multivalent interactions.

On the contrary, particles carrying receptor-cleaving ligands can crawl across functionalized surfaces without detaching (Fig.~\ref{Fig:Motility}). At short timescales, their motion is not Brownian but persistent  \cite{sakai_influenza_2017,vahey_influenza_2019,agarwal_kinetics_2024,stevens_sliding_2023}. Such a Brownian-ratchet movement arises from NA enzymes that deplete the surface beneath the virion from receptors and push the particle to move towards regions richer in SIA to maximize the number of bonds. Different positioning of the NA enzymes on the surface of the virion results in different types of motion  (Fig.~\ref{Fig:Motility}b). For instance,  filamentous virions with polarized coatings, with NA ligands concentrated at one of the two tips of the particle, preferentially move along their axial direction \cite{vahey_influenza_2019,stevens_sliding_2023}.
Instead, uniform distribution of the ligands result in filamentous virions that move preferentially on the side, orthogonally to their axial direction \cite{stevens_sliding_2023}. Existing numerical studies \cite{vahey_influenza_2019,stevens_sliding_2023} simplify the geometry of the virions into a 2D capsule (representing the contact area between a spherocylinder and a plane) moving parallel to the cell surface. It will be important to consider more realistic 3D models that also include spherical particles, as anticipated in Fig.~\ref{Fig:Motility}.

Persistent types of motion have also been reported in systems featuring the so-called burnt-bridge mechanism, including Influenza C systems \cite{sakai_unique_2017,ziebert_how_2021,ziebert_how_2021} or synthetic molecular walkers \cite{perl2011gradient,bazrafshan_tunable_2020,olah_superdiffusive_2013}.
% suggested references: [korosec_lawnmower_2021] (Korosec et al. 2021, arXiv -- lawnmower burnt-bridge molecular motor, relevant to synthetic molecular walker context)
In these types of systems, a single type of ligand binds and cleaves the receptors at the surface. This results in a persistent type of motion similar to what is found in IAV systems with NA enzymes uniformly distributed at the virion surface \cite{sakai_influenza_2017,ziebert_how_2021,stevens_sliding_2023}.

\subsection{IAV Mobility through Mucus}\label{Sec:Mucus}

The movement of IAV through respiratory mucus represents a more complex situation than its motion across receptor-functionalized two-dimensional surfaces of host cells. Mucus is a three-dimensional, heterogeneous polymer network in which virions experience both steric confinement by the mucin mesh and transient adhesive interactions with sialylated glycans. Consequently, IAV mobility reflects an interplay between mucus microstructure, which imposes size-dependent steric constraints on particles relative to a characteristic mesh size of approximately 100--500\,nm, HA-mediated binding, NA-mediated receptor cleavage, and intrinsic virion properties such as size, morphology, and the spatial organization of HA and NA.

Early experimental studies provided evidence that NA activity facilitates IAV penetration through mucus. Single-particle tracking of swine influenza virus in porcine respiratory mucus revealed a heterogeneous population in which approximately 70\% of particles were trapped and the remaining fraction remained mobile. Inhibition of NA with zanamivir reduced both microscopic particle mobility and macroscopic mucus penetration, whereas addition of exogenous neuraminidase enhanced penetration, supporting a model in which mucin-associated sialic acids act as decoy receptors that can transiently immobilize virions \cite{yang2014beneficiary}. The spatial organization of HA and NA may further influence this process: for instance, in filamentous IAV particles, polarization of NA toward one end of the virion generates persistent motion away from the NA-rich pole on receptor-coated surfaces, consistent with the Brownian-ratchet mechanism described in the previous section, in which local receptor destruction biases subsequent movement toward regions containing uncleaved receptors \cite{vahey_influenza_2019}. Such a mechanism has been proposed to facilitate directional movement through sialylated mucus.

However, studies using human airway mucus indicate that receptor cleavage may not be the sole, or even the dominant, determinant of IAV mobility. Multiple-particle tracking experiments showed that IAV diffusion strongly correlates with the microstructure of the mucus gel. Virus-sized, muco-inert nanoparticles exhibited similar restrictions in mobility, and inhibition of viral NA produced only minor changes in IAV diffusion in the tested human mucus samples \cite{kaler_influenza_2022}. More recent experiments using mucus produced by differentiated airway epithelial cultures similarly found that mucus with a larger fraction of virus-sized pores more effectively restricted IAV mobility. Depletion of mucin-associated sialic acids did not substantially increase virus diffusion, and synthetic PEG hydrogels lacking sialic acid nevertheless provided an effective physical barrier against IAV when their network dimensions approached those of airway mucus \cite{kaler_mucus_2025}. These observations indicate that steric obstruction by the mucus network can make a substantial contribution to viral trapping independently of HA--SIA interactions.

Virion morphology may provide an additional determinant of mucus penetration. Filamentous IAV particles have recently been reported to penetrate mucus more efficiently than predominantly spherical virus populations. These particles contained relatively more NA and displayed markedly higher NA activity, suggesting that both virion geometry and the HA--NA balance may influence transport through the mucosal barrier \cite{liu2026filamentous}. Together, these findings indicate that IAV mobility in mucus cannot be described solely as receptor-mediated crawling. Instead, transport emerges from the coupling of multivalent HA--SIA interactions and NA-mediated receptor destruction with the three-dimensional architecture and dynamics of the mucus network. Developing realistic three-dimensional reaction--diffusion models that incorporate mucus pore structure, mucin mobility, heterogeneous receptor distributions, and virion morphology will therefore be important for connecting molecular HA--NA kinetics to experimentally observed mucus penetration.

\section{Conclusions and Perspectives}\label{Sec:Perspective}

Multivalent interactions provide a unifying framework through which to understand how influenza A virions achieve stable yet dynamic engagement with sialylated host environments.
This work employs multivalent models (Sec.~\ref{Sec:Multivalent}) to rationalize the dynamics (Sec.~\ref{Sec:Lim_Multi}) and targeting  (Sec.~\ref{Sec:Nanom}) of IAV virions.

The collective action of numerous weak haemagglutinin–sialic acid bonds provides a mechanistic explanation for the high avidity of virion attachment and the ability of virions to sense receptor density.
Multivalent models (Sec.~\ref{Sec:Multivalent}) estimate avidity using statistical mechanics. 
Future work will need to further develop such quantitative theories, accounting for non-specific interactions, including steric, hydrophobic, and electrostatic contributions. 
These terms are non-negligible and may result in qualitative alteration of the phase behavior, such as re-entrant adsorption curves \cite{liu2020combinatorial}, with cell-virion binding occurring only within a finite range of receptor concentrations. 
Current multivalent theories do not account for the reaction kinetics (the fact that the equilibration of the number of bonds is not instantaneous), which could also distort the equilibrium phase behaviors \cite{lanfranco_kinetics_2019,bachmann_bond_2016}. In addition, the spatial organisation of HA and NA and the pleomorphic nature of IAV particles may influence the relationship between molecular binding parameters and the collective behaviour of the virion. Therefore, a central challenge is to connect quantitative single-molecule parameters with whole-virion behaviour in realistic extracellular environments.

In the development of antiviral strategies (Sec.~\ref{Sec:Nanom}), lack of accurate models has been partly compensated for by designs mimicking key parameters of the cell surface, such as the density of SIA groups \cite{reiter-scherer_force_2019}. It will be interesting to generalize these types of approaches, ideally integrating theoretical and numerical models. This may open new avenues to inform multivalent models using experimental results. For instance, modern microscopic techniques allow charting NA and HA position at the single-molecule level (see Fig.~\ref{Fig:General_Model}). It should then be possible to compare {\em in-silico} (Sec.~\ref{Sec:Motility}) and {\em in-vitro} trajectories of virions crawling across reconstructed cell surfaces \cite{vahey_influenza_2019}. Such comparisons may open the avenue to estimate association and dissociation rates, the former including configurational terms that are not accounted for in existing experimental measurements of the reaction rates (Sec.~\ref{Sec:LR} and Fig.~\ref{Fig:Bond Interaction}). Combining these measurements with controlled receptor-presenting surfaces and structurally defined antiviral scaffolds should make it possible to test how receptor density, ligand arrangement, and binding kinetics determine virion attachment and inhibition.

When studying virion dynamics, it is pivotal to consider the effect of NA catalytic activity. By cleaving SIA receptors, NA can drive the system out of equilibrium, enabling virion detachment and surface motility. NA activity has been modeled almost exclusively using numerical simulations. It will therefore be important to develop theories capable of distilling key aspects of the NA-driven dynamics. For instance, in Fig.~\ref{Fig:Nb_t_Peak}, we show how the number of bonds (in systems with highly mobile receptors) fluctuates around a slowly changing baseline. This observation suggests a separation of timescales, with the number of bonds relaxing on shorter timescales than the change in receptor concentration. This scenario could be captured by multivalent theories with time-dependent receptor concentrations. Such approaches would provide a basis for describing the transition from equilibrium-like receptor binding to nonequilibrium virion motion and for quantifying the functional balance between HA-mediated attachment and NA-mediated receptor cleavage. Integrating single-particle tracking with these theoretical descriptions could further reveal how molecular reaction rates translate into virion trajectories and how this relationship changes in the presence of antiviral compounds.

Finally, current models of cell-virion interactions consider simplified cell surfaces, as typically found in experiments using supported lipid bilayers. Future advances will need to consider the effect of the glycocalyx on the motion of the virion and, especially, its ability to navigate through the mucus layers. The heterogeneous, three-dimensional architecture of the glycocalyx and mucus may affect receptor accessibility, steric constraints and the transport of virions towards susceptible cells. The role of NA activity in mucus is still debated as discussed in the previous section
\cite{cohen_influenza_2013,yang2014beneficiary,wang2016influenza,kaler_influenza_2022,kaler_mucus_2025,liu2026filamentous}. It will be important to determine how the interplay between multivalent attachment, receptor cleavage and steric interactions governs the ability of virions to move through the host mucus. A quantitative understanding of the dynamics of virion particles in the extracellular medium will allow predicting virion trafficking from binding constants and testing the functional role of HA-NA synergy under increasingly realistic conditions 
(Sec.~\ref{Sec:HA-NA_synergy}). Ultimately, an experimentally grounded, nonequilibrium theory of IAV transport, adhesion and inhibition could connect molecular binding parameters to whole-virion behaviour and guide the development of antiviral strategies combining high-avidity binding, NA inhibition and steric shielding.

\section*{Acknowledgments}

SB and LK are supported by funding from Novo Nordisk Foundation (Grant No: NNF23SA0088060). DCL is grateful for financial support from the federal ministry of education and research (BMBF) funded project MucPep (FKZ: 13XP511). AH and DCL acknowledge the financial support of the Collaborative Research Center “Dynamic Hydrogels at Biological Interfaces” (CRC 1449) funded by the Deutsche Forschungsgemeinschaft (DFG, German Research Foundation)--Project ID 431232613-SFB 1449.
TK and BMM are supported by a PDR grant of the FRS-FNRS (Grant No. T021024F).

%\printbibliography
%\bibliographystyle{unsrt}
%\bibliography{references}

\end{document}